\documentclass[preprintnumbers,twocolumn,showpacs,amsmath,amssymb]{revtex4}

\usepackage{epsfig}

\usepackage{graphicx}

\usepackage{dcolumn}

\usepackage{bm}

\newcommand{\be}{\begin{equation}}

\newcommand{\ee}{\end{equation}}

\newcommand{\nn}{\nonumber}

\newcommand{\bea}{\begin{eqnarray}}

\newcommand{\eea}{\end{eqnarray}}

\newcommand{\bfig}{\begin{figure}}

\newcommand{\efig}{\end{figure}}

\newcommand{\bc}{\begin{center}}

\newcommand{\ec}{\end{center}}

\def\ycut{y_{{\mathrm{cut}}}}

\def\as{\alpha_s}

\def\asmz{\alpha_s(M_Z)}

\def\epm{e^+e^-}

\def\lapprox{\lower .7ex\hbox{$\;\stackrel{\textstyle <}{\sim}\;$}}

\def\gapprox{\lower .7ex\hbox{$\;\stackrel{\textstyle >}{\sim}\;$}}

\begin{document}
\preprint{ZU-TH 15/09, IPPP/09/87, ETH-IPP-2009-11} 
\title{Precise determination of the strong coupling constant at NNLO in QCD from
the three-jet rate in electron--positron annihilation at LEP}
\author{G.\ Dissertori$^a$, 
A.\ Gehrmann-De Ridder$^b$, T.\ Gehrmann$^c$, E.W.N.\ Glover$^d$,
G.\ Heinrich$^d$, H.\ Stenzel$^e$}
 \affiliation{$^a$ Institute for Particle Physics, ETH Zurich, CH-8093 Zurich, Switzerland\\
 $^b$ Institute for Theoretical Physics, ETH Zurich, CH-8093 Zurich, Switzerland\\
$^c$ Institut f\"ur Theoretische Physik,
Universit\"at Z\"urich, CH-8057 Z\"urich, Switzerland\\
$^d$ Institute for Particle Physics Phenomenology, 
        Department of Physics,
        University of Durham, Durham, DH1 3LE, UK\\
$^e$ II. Physikalisches Institut, Justus-Liebig Universit\"at Giessen, 
D-35392 Giessen, Germany      
        }
\date{\today}

\begin{abstract}
We present the first determination of the strong coupling constant from
the three-jet rate in $\epm$ annihilation at LEP, based on a next-to-next-to-leading order (NNLO) perturbative QCD prediction.  More precisely, we extract $\asmz$ by fitting
perturbative QCD predictions at ${\cal O}(\alpha_s^3)$ to data from the 
ALEPH experiment at LEP. Over a large range of the jet-resolution
parameter $\ycut$ this observable is characterised by small non-perturbative corrections
and an excellent stability under renormalisation scale variation. We find $\asmz = 0.1175 \pm 0.0020\, (\mathrm{exp}) \pm 0.0015\, (\mathrm{theo})$,
which is  more accurate than the values of $\asmz$ from $e^+e^-$ event shape data currently used in the world average.

\end{abstract}
\maketitle

Jet observables in  electron--positron
annihilation play an outstanding role in 
studying the dynamics of 
the strong interactions \cite{Ellis:1976uc}, 
described by the theory of Quantum Chromodynamics (QCD,~\cite{qcd}).  
In particular, jet rates  and related event-shape observables have been extensively used
for the determination  of the QCD coupling constant $\as$
(see~\cite{reviews,bethke} for a review), mostly based on data obtained
at the $\epm$ colliders PETRA, LEP and SLC at centre-of-mass energies from
$14$ to $209$ GeV.
Jets are defined using 
a jet algorithm, which describes how to recombine the particles in an event to 
form the jets. A jet algorithm consists of  two ingredients: a 
distance measure and a recombination procedure.
The distance measure 
is computed for each pair of particles to select the pair with the smallest 
separation in momentum space. If the separation is below a pre-defined resolution parameter $\ycut$,
the pair are combined 
according to the recombination procedure.
The JADE algorithm~\cite{jade} uses the pair
invariant mass as distance measure. Several improved jet algorithms  
have been proposed for $e^+e^-$ collisions: Durham~\cite{durham},
Geneva~\cite{geneva} and Cambridge~\cite{cambridge}. 
The Durham algorithm has been the most widely used by experiments at 
LEP~\cite{aleph,opal,delphi,l3} and SLD~\cite{sld}, as well as 
in the reanalysis of earlier data at lower energies from 
JADE~\cite{jadenew}.

The Durham jet algorithm clusters utilises the distance measure
\begin{equation}
y_{ij,D} = \frac{2 \, {\rm min} (E_i^2,E_j^2) (1-\cos \theta_{ij})}
{E_{{\rm vis}}^2}
\end{equation}
for each pair ($i,j$) of particles, $E_{{\rm vis}}$ 
denotes the energy sum of all particles in the final state.
The pair with the lowest 
$y_{ij,D}$ is replaced by a pseudo-particle whose four-momentum is 
given by the 
sum of the four-momenta of particles $i$ and $j$ ('$E$' recombination 
scheme). This procedure is repeated as long as pairs with invariant 
mass below the predefined resolution parameter
$y_{ij,D}<\ycut$  are found. Once the clustering is terminated, the 
remaining (pseudo-)particles are the jets. 
In experimental jet measurements, 
one studies the jet rates, i.e. jet cross sections normalised to 
the total hadronic cross section, as function of the jet-resolution 
parameter $\ycut$.

The theoretical prediction of jet cross sections 
is made within perturbative QCD, where the same jet algorithm is applied 
to the final state partons. The QCD description of jet production 
is either based on a fixed-order calculation or a parton shower.
The fixed order approach
uses exact parton-level
matrix elements including higher order corrections where available and/or 
analytical resummation of large logarithmic corrections 
for a given jet multiplicity. On the other hand, the parton shower starts 
with the leading-order matrix element for two-jet production   
and generates higher multiplicities in an iterative manner, thereby accounting only 
for the leading logarithmic terms from parton-level processes with 
higher multiplicity. In multi-purpose event generator 
programs~\cite{herwig,ariadne,pythia}, such parton showers are
complemented by phenomenological models which describe
the transition from partons to hadrons. These programs provide
a satisfactory description of multi-jet production rates but, since 
they generally contain many tunable phenomenological parameters, their 
predictive power is limited. Nevertheless, in order to compare parton level predictions with experimental hadronic data, these event generators are vital to estimate the effects due to hadronisation and resonance decays.

Until recently, fixed-order calculations were available 
up to next-to-next-to-leading order (NNLO) for two 
jets~\cite{babis2j,our2j,weinzierl2j} and up to next-to-leading order 
(NLO) for three~\cite{ERT,kunszt,event} and 
four jets~\cite{dixonsigner4j,nagy4j,cullen4j,weinzierl4j}. 
For five and more jets, only leading order calculations are 
available~\cite{tree5p,moretti6j,amegic}. For jets involving massive quarks,
NLO results are available for three-jet final states~\cite{quarkmass}. 
 The recent calculations of the $\alpha_s^3$ corrections (NNLO) for three-jet production
\cite{ourevent,our3j,Weinzierl:2008iv,Weinzierl:2009ms}
have already led to precise $\as$ determinations 
\cite{Dissertori:2007xa,becherschwartz,davisonwebber,Dissertori:2009ik,Bethke:2008hf}, using event-shape observables
 measured by ALEPH and JADE. However, some of the
 event-shape variables still suffer from 
a poor convergence of the perturbative expansion even at NNLO.
 Furthermore, the usage of event generators, which have been tuned to LEP data, 
 for the determination of the hadronisation corrections may lead to
 a bias in the $\as$ measurements for some of the event shapes \cite{Dissertori:2009ik}. 
A comparison of different variables showed that jet broadening variables are most affected by 
missing higher orders and a potential hadronisation bias, while the differential two-jet rate $Y_3$ is 
most robust against these effects, and strongly motivates the present study of the three-jet rate.

In this letter we describe a determination of the strong coupling constant
from the three-jet rate measured by ALEPH \cite{ALEPH-qcdpaper} at LEP. We use the NNLO
predictions as  presented in \cite{our3j}. There it was shown that:
(i) For large values of $\ycut$, $\ycut > 10^{-2}$, the NNLO corrections 
turn out to be very small, while they become substantial for medium and 
low values of $\ycut$; 
(ii) The maximum of the jet rate is shifted towards 
higher values of $\ycut$ compared to NLO and is in better 
agreement with the experimental observations; 
(iii) The theoretical uncertainty is lowered considerably compared to NLO, 
especially in the region $10^{-1}> \ycut > 10^{-2}$ relevant 
for precision phenomenology where the theory error is below two per-cent relative uncertainty; 
(iv) Finally,  in this $\ycut$ region the parton level predictions at NNLO are
already very close to the experimental measurements, indicating the need
for only small hadronisation corrections.

These findings motivate a dedicated analysis of the three-jet rate, leading to a
precise measurement of $\as$. Our analysis closely follows the procedure
described in \cite{Dissertori:2007xa,Dissertori:2009ik}. 
The ALEPH data \cite{ALEPH-qcdpaper} at LEP
are based on the reconstructed momenta and
energies of charged and neutral particles. The measurements have
been corrected for detector effects, i.e.\ the final distributions
correspond to the so-called particle (or hadron) level, 
and for initial state photonic radiation. 
In the simulation of the detector response to particles, a bias is introduced 
by the choice of the physics event generator. This leads to a systematic 
uncertainty on the three-jet rate of about $1.5\%$ for
the relevant $\ycut$ range. Further experimental systematic effects 
are estimated by a variation of the track- and event-selection cuts as advocated 
in \cite{ALEPH-qcdpaper}, giving an additional small systematic uncertainty of about $1\%$. 

We construct the perturbative expansion up to ${\cal O}(\alpha_s^3)$ 
as described in \cite{Dissertori:2009ik}, with the coefficients obtained from
\cite{our3j}. These are valid for massless quarks. 
We take into account bottom mass effects up to NLO \cite{quarkmass},
for a pole b-quark mass of $M_{\rm b} = 4.5$ GeV. The latter is varied
by $\pm 0.5$ GeV in order to estimate the impact of the b-quark mass
uncertainty on the value of the strong coupling. 
For the normalisation to the total hadronic cross section $\sigma_{{\rm had}}$
we follow the procedure adopted in \cite{Dissertori:2009ik}, 
which is based on a N$^3$LO calculation (${\cal O}(\alpha_s^3)$ in QCD) 
for  $\sigma_{\mathrm{had}}$ \cite{kuhnrev},
including mass corrections for the b-quark up to
${\cal O}(\alpha_s)$ and the leading
mass terms to ${\cal O}(\alpha_s^2)$. Weak corrections to the three-jet 
rate were computed very recently~\cite{Denner:2009gx}.
They are at the one per-mille
level for $Q=M_Z$ and are neglected here.

The nominal value for the renormalisation scale $x_\mu = \mu/Q$ is unity. 
It is varied between $0.5 < x_\mu < 2$ in order to assess the systematic
uncertainty related to yet unknown higher order corrections.
No attempt is made to combine the NNLO predictions with resummation
calculations. At present, the resummation of the three-jet rate~\cite{durham} 
is only fully consistent at leading logarithmic level~\cite{resumy3}, 
and resummation effects only become numerically relevant over fixed-order NNLO  
 for $\ln\ycut\lapprox -4.5$ (as can be seen from the $Y_3$ transition 
parameter distribution~\cite{gionata}), 
 which is  below our region of interest.

In order to compare the perturbative parton level thoretical prediction with the hadronic data, it is necessary to apply a correction for hadronisation and resonance decays.  This bin-by-bin correction is
computed with the PYTHIA \cite{pythia}, HERWIG  \cite{herwig} and ARIADNE \cite{ariadne} Monte Carlo generators, all tuned to global hadronic observables at $M_Z$
\cite{aleph_mega}. The parton level is defined by the quarks and
gluons present at the end of the parton shower in PYTHIA and HERWIG
and the partons resulting from the colour dipole radiation in
ARIADNE. Our central values for the strong coupling constant are obtained
with hadronisation corrections from PYTHIA, which are at the level of $5\%$. 
We define the systematic uncertainty on $\asmz$ due to these hadronisation
corrections as the biggest deviation observed when using  any of the other generators.
Motivated by the observations in \cite{Dissertori:2009ik}, we verified that the
shapes of the 
Monte Carlo parton level predictions are in fair agreement with those at NNLO,
for reasonable choices of the strong coupling. Furthermore, the ratios of these
predictions are relatively flat over the relevant $\ycut$ range, giving further confidence
in the reliability of the hadronisation corrections.

\begin{figure}[h]
\epsfig{file=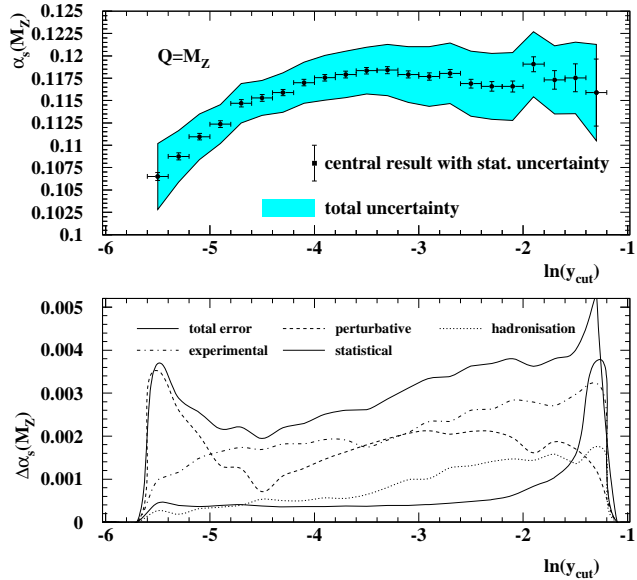,width=8.5cm}
\caption{Determinations of $\asmz$ from the three-jet rate, measured by ALEPH at the $Z$ peak, for several
           values of the jet-resolution parameter $\ycut$. The error bars show the statistical uncertainty,
            whereas the shaded band indicates the total error, including the systematic uncertainty. The various
             contributions to the latter are displayed in the lower plot. \label{fig:asresults}}
\end{figure}

The corrected ALEPH measurements for the three-jet rate are
compared to the theoretical calculation at particle level. Values for $\asmz$
are obtained by a least-squares fit, performed separately for each $\ycut$ value
in the range  listed in Table \ref{tab:fitresult} (for the data at the $Z$ peak),
together with the uncertainties as described above. These results are
also displayed in Fig.\ \ref{fig:asresults}. We observe a nice stability of the results,
within their total uncertainties, down to resolution parameters of $\ln\ycut \approx -4.5$.
Beyond that value we find a fall-off of $\asmz$, most likely  related to the 
onset of large logarithmic corrections from higher perturbative orders,
which are not accounted for in our perturbative prediction. 

As final result we quote our measurement for $\ycut = 0.02$, which 
represents an optimal compromise between minimal systematic uncertainty and stability. We find
\be
\label{eq:ourresult}
 \nn
  \asmz = 0.1175 \pm 0.0020\, (\mathrm{exp}) \pm 0.0015\, (\mathrm{theo})  
\ee
where the first uncertainty includes (in quadrature) the contributions from statistics, detector corrections
and experimental selection cuts, and the second error is the quadratic sum of b-quark
mass and renormalisation scale uncertainties (cf.\ Table \ref{tab:fitresult}). We also performed similar measurements 
for the LEP2 energies between 133 and 206 GeV, where we find consistent 
values for $\asmz$, but with considerably larger statistical uncertainties.
Combining the errors in quadrature, yields $\asmz = 0.1175 \pm 0.0025$
which is in excellent agreement with the latest world average value \cite{bethke}
of $\asmz = 0.1184 \pm 0.0007$
that is based on a number of measurements from $\tau$-decay, lattice gauge theory, Upsilon decay, DIS and $e^+e^-$ data.
As expected, our
theoretical uncertainty is smaller than that obtained from fits of event-shape distributions,
and even smaller than the experimental error, which is dominated by the model-dependence of
the detector corrections. Our result is also more precise than the two extractions of $\alpha_s$ from $e^+e^-$ event-shape data \cite{Bethke:2008hf,Dissertori:2009ik} currently used in the world average \cite{bethke}.

In this letter we reported on the first determination of the strong coupling constant from
the three-jet rate in $\epm$ annihilation at LEP, based on a NNLO perturbative QCD prediction.
We find a precise value of $\asmz$ with an uncertainty of $2\%$, consistent with the world average.
This verifies the expectations that the three-jet rate is an excellent observable for this kind
of analysis, thanks to the good behaviour of its perturbative and non-perturbative contributions over
a sizable range of jet-resolution parameters.
\begin{table}[b]
{\small
\hfill{}
{\scriptsize
\begin{tabular}{c|cccccccc}
$\ln(\ycut)$ &  $\asmz$ & stat. & det. & exp. & had. & mass & pert. & total  \\ \hline 
-5.1  &  0.1110  &  0.0004  &  0.0013  &  0.0008  &  0.0003  &  0.0004  &  0.0020  &  0.0025  \\
-4.9  &  0.1124  &  0.0004  &  0.0015  &  0.0007  &  0.0003  &  0.0003  &  0.0013  &  0.0022  \\
-4.7  &  0.1147  &  0.0004  &  0.0015  &  0.0008  &  0.0004  &  0.0003  &  0.0012  &  0.0022  \\
-4.5  &  0.1153  &  0.0004  &  0.0015  &  0.0008  &  0.0005  &  0.0003  &  0.0006  &  0.0019  \\
-4.3  &  0.1159  &  0.0004  &  0.0016  &  0.0009  &  0.0005  &  0.0003  &  0.0010  &  0.0022  \\
-4.1  &  0.1170  &  0.0004  &  0.0016  &  0.0009  &  0.0005  &  0.0003  &  0.0012  &  0.0023  \\
-3.9  &  0.1175  &  0.0004  &  0.0016  &  0.0011  &  0.0006  &  0.0002  &  0.0014  &  0.0025  \\
-3.7  &  0.1179  &  0.0004  &  0.0016  &  0.0011  &  0.0006  &  0.0002  &  0.0016  &  0.0026  \\
-3.5  &  0.1183  &  0.0004  &  0.0015  &  0.0009  &  0.0006  &  0.0002  &  0.0018  &  0.0026  \\
-3.3  &  0.1184  &  0.0004  &  0.0015  &  0.0011  &  0.0008  &  0.0002  &  0.0019  &  0.0029  \\
-3.1  &  0.1179  &  0.0004  &  0.0016  &  0.0013  &  0.0010  &  0.0002  &  0.0021  &  0.0031  \\
-2.9  &  0.1177  &  0.0004  &  0.0019  &  0.0013  &  0.0010  &  0.0002  &  0.0021  &  0.0033  \\
-2.7  &  0.1180  &  0.0004  &  0.0020  &  0.0013  &  0.0013  &  0.0001  &  0.0020  &  0.0034  \\
-2.5  &  0.1169  &  0.0005  &  0.0021  &  0.0015  &  0.0013  &  0.0001  &  0.0021  &  0.0036  \\
-2.3  &  0.1166  &  0.0005  &  0.0019  &  0.0018  &  0.0014  &  0.0001  &  0.0021  &  0.0037  \\
-2.1  &  0.1166  &  0.0006  &  0.0020  &  0.0020  &  0.0015  &  0.0001  &  0.0020  &  0.0038  \\
-1.9  &  0.1191  &  0.0008  &  0.0021  &  0.0019  &  0.0014  &  0.0002  &  0.0016  &  0.0036  \\
-1.7  &  0.1173  &  0.0010  &  0.0015  &  0.0023  &  0.0016  &  0.0001  &  0.0019  &  0.0038  \\
-1.5  &  0.1175  &  0.0016  &  0.0005  &  0.0029  &  0.0014  &  0.0001  &  0.0017  &  0.0040  \\
-1.3  &  0.1159  &  0.0037  &  0.0014  &  0.0029  &  0.0018  &  0.0004  &  0.0011  &  0.0054  \\ 
\hline
\end{tabular}}
\hfill{}
\caption[3-jet fits]{\label{tab:fitresult}{Results of $\asmz$ extracted from 
the three-jet rate measured by ALEPH at LEP1. The uncertainty contributions are given for 
the statistical error (stat.), the uncertainty related to the choice of the generator for the simulation of the 
detector response (det.), the quadratic sum of all other experimental systematic uncertainties arising 
from track and event selection cut variations (exp.), the hadronisation uncertainty obtained 
by the maximum difference between either PYTHIA, HERWIG or ARIADNE (had.), the 
uncertainty on the b-quark mass correction procedure (mass) and the uncertainty for
missing higher orders (pert.) estimated by a variation of the renormalisation scale. }}}
\end{table}
\\ 
{\noindent \bf Acknowledgements:} 
This research was supported in part by the Swiss National Science Foundation
(SNF) under  contracts PP0022-118864 and
200020-126691,
 by the UK Science and Technology Facilities Council, by the European Commission's Marie-Curie Research Training Network under contract
MRTN-CT-2006-035505 ``Tools and Precision Calculations for Physics Discoveries
at Colliders'' and by the German Helmholtz Alliance ``Physics at the Terascale''.
EWNG gratefully acknowledges the support of the Wolfson Foundation and
the Royal Society.  


\end{document}